# Some Applications of the Percolation Theory: Brief Review of the Century Beginning

Alexander Herega

*Department of Computer Systems, Odessa National Academy of Food Technologies, Odessa, 65039, Ukraine*

**Abstract:** The review is a brief description of the state of problems in percolation theory and their numerous applications, which are analyzed on base of interesting papers published in the last 15-20 years. At the submitted papers are studied both the cluster system of the physical body and its impact on the object in general, and adequate mathematical tools for description of critical phenomena too. Of special interest are the data, first, the point of phase transition of certain of percolation system is not really a point, but it is a critical interval, and second, in vicinity of percolation threshold observed many different infinite clusters instead of one infinite cluster that appears in traditional consideration.

## 1. Introduction

The percolation theory is a section of probability theory, which has multiple applications in natural and engineering sciences [1-4]. The percolation theory, deals with the special features of the appearance and evolution as well as the properties of clusters arising in stochastic processes.

Stable interest of specialists in percolation structures that is observed in the last decades can be easily explained by the obvious significance of the study of critical phenomena. The generated percolation structures cardinally modify the material: takes place a structural phase transition, the correlation length is jumping, the object symmetry and other parameters are changing, thus leading to changes in physical-chemical and mechanical characteristics of physical bodies. Such clusters substantially modify the conductivity processes, affect the kinetics of chemical reactions, define the mechanical strength and corrosion resistance, and lead to abnormal diffusion and other phenomena. Therefore, in percolation studies one usually investigates both the cluster system of the physical body and its impact on the object in general simultaneously.

The percolation theory represents adequate mathematical tools and the possibility of physical description of phenomena, caused by the appearance of linked (quasi-linked) domains having typical body dimensions, i.e., percolation clusters of arbitrary nature: phases, defects, interfaces, etc.

## 2. Variations of "Standart" Problems

The internal and external surfaces of a percolation cluster, as well as the total surface of the entire percolation system, in [5] are investigated numerically and analytically. Numerical simulation is carried out using the Monte Carlo method for problems of percolation over lattice sites and bonds on square and simple cubic lattices. Analytic expressions derived by using the probabilistic approach describe the behavior of such surfaces to a high degree of accuracy. It is shown that both the external and total surface areas of a percolation cluster, as well as the total area of the surface of the entire percolation system, have a peak

---

**Corresponding author:** Alexander Herega, professor, chief of Department of Computer Systems, research fields: material science, percolation theory, fractal clusters, dynamical systems, deterministic chaos, computer simulation and modeling of physical phenomenon.

for a certain (different in the general case) fraction of occupied sites (in the site problem) or bonds (in the bond problem). Two examples of technological processes (current generation in a fuel cell and self-propagating high-temperature synthesis in heterogeneous condensed systems) in which the surface of a percolation cluster plays a significant role are discussed [5].

The bond-percolation process is studied in [6] on periodic planar random lattices and their duals. Authors determined the thresholds and critical exponents of the percolation transition; the scaling functions of the percolating probability, the existence probability of the appearance of percolating clusters, and the mean cluster size are also calculated. The simulation result of the percolation threshold is $p_c = 0.3333 \pm 0.0001$ for planar random lattices, and $0.6670 \pm 0.0001$ for the duals of planar random lattices. Authors conjecture that the exact value of $p_c$ is 1/3 for a planar random lattice and 2/3 for the dual of a planar random lattice. By considering possible errors, the results of critical exponents agree with the values given by the universality hypothesis. By properly adjusting the metric factors on random lattices and their duals, and demonstrate explicitly that the idea of a universal scaling function with no universal metric factors in the finite-size scaling theory can be extended to random lattices and their duals for the existence probability, the percolating probability, and the mean cluster size.

Authors [7] investigate site percolation on a weighted planar stochastic lattice, which is a multifractal and whose dual is a scale-free network. Percolation is typically characterized by a threshold value $p_c$ at which a transition occurs and by a set of critical exponents β, γ, ν, which describe the critical behavior of the percolation probability P (p), mean cluster, size S (p), and the correlation length ξ. Besides, the exponent τ characterizes the cluster size distribution function $n_s(p_c)$ and the fractal dimension $d_f$ characterizes the spanning cluster. In paper numerically obtain the value of $p_c$ and of all the exponents. These results suggest that the percolation on weighted planar stochastic lattice belong to a separate universality class than on all other planar lattices.

In [8] studied a model of semi-directed percolation on finite strips of the square and triangular lattices. Using the transfer-matrix method, and phenomenological renormalization group approach, authors obtains good numerical estimates for critical probabilities and correlation lengths critical exponents. Our results confirm the conjecture that semi-directed percolation belongs to the universality class of the usual fully directed percolation model.

## 3. Interpretation of Percolation in Terms of Infinity Computations

In [9] a number of traditional models related to the percolation theory have been considered by means of new computational methodology that does not use Cantor's ideas and describes infinite and infinitesimal numbers in accordance with the principle "The part is less than the whole". It gives a possibility to work with finite, infinite, and infinitesimal quantities numerically by using the so-called Infinity Computer – A new kind of a computer introduced in [10].

Site percolation and gradient percolation have been studied by applying the new computational tools. It has been established that in an infinite system the phase transition point is not really a point as with respect of traditional approach. In light of new arithmetic it appears as a critical interval, rather than a critical point. Depending on "microscope" in [9] use this interval could be regarded as finite, infinite and infinitesimal short interval. Using new approach, authors observed that in vicinity of percolation threshold exist many different infinite clusters instead of one infinite cluster that appears in traditional consideration.

## 4. The "Explosive" Percolation

The basic notion of percolation in physics assumes the emergence of an "infinite" percolation cluster in a large disordered system when the density of connections exceeds some critical value (except, for example, for percolation with a threshold at zero [11]). Until recently, the percolation phase transitions were believed to be continuous, however, authors [12] was reported about the new so-called "explosive" percolation problem, which is a remarkably different discontinuous phase transition. A specific optimization process establishes each of bonds in this problem. It is important the optimization rules can both delay and accelerate percolation in such models, which now known as Achlioptas processes [13, 14].

Employing analytical and numerical calculations, authors [15] showed the "explosive" percolation transition is continuous though with a uniquely small critical exponent of the percolation cluster size. These transitions provide a new class of critical phenomena in irreversible systems and processes [15].

The Gaussian model of discontinuous percolation is numerically investigated in three dimensions, disclosing a discontinuous transition [16]. For the simple cubic lattice, in the thermodynamic limit authors report a finite jump of the order parameter $J = 0.415 \pm 0.005$. The largest cluster at the threshold is compact, but its external perimeter is fractal with fractal dimension $d = 2.5 \pm 0.2$. The study is extended to hypercube lattices up to six dimensions and to the mean-field limit (infinite dimension). In [16] find that, in all considered dimensions, the percolation transition is discontinuous. The value of the jump in the order parameter, the maximum of the second moment, and the percolation threshold are analyzed, revealing interesting features of the transition and corroborating its discontinuous nature in all considered dimensions. In the paper also shown that the fractal dimension of the external perimeter, for any dimension, is consistent with the one from bridge percolation and establish a lower bound for the percolation threshold of discontinuous models with a finite number of clusters at the threshold.

## 5. Percolation on the Fractal Objects

In [17] proposed modification of the Sierpinski carpet consists in assuming that the cells having a common edge or vertex are connected, and refer to this analog of the known fractal as a Sierpinski carpet with hybrid ramification. Apparently, the modification of rules defining the connectedness leads to a change in percolation parameters of an infinite cluster of carpet cells. Authors determine the probability $p'$ that a cell belongs to a percolation cluster on the carpet, i.e., the probability that there is a "flow" through the squares composing it, each belonging to an infinite cluster with the probability $p$. Since the renormalization group transformation [18] in our case reflect should the connectedness, the number of suitable combinations in the arrangement of squares in the cell is less than the combinatorial one. Therefore, the renormalization-group transformation for a carpet with hybrid ramification takes the form:

$$p' = R(p) = p^8 + 8p^7(1-p) + 27p^6(1-p)^2 + \\ + 44p^5(1-p)^3 + 38p^4(1-p)^4 + 8p^3(1-p)^5 \quad (1)$$

with a nontrivial stationary point $p_c = 0.5093$, which defines the percolation threshold.

The index of the correlation length of the percolation system can be found from the relation $v = \ln b / \ln \lambda = 1.801$, where $b = 3$ is the number of squares along the cell side and

$$\lambda = (dR/dp)\big|_{p=p_c} \quad (2)$$

The critical exponent of the order parameter $\beta$ is determined from the equality $D = d - \beta/v$, where the dimension $D$ of the percolation cluster can be approximated by that of the Sierpinski carpet; for the spatial dimension $d = 2$, $\beta = 0.193$. (To verify the obtained values: in the case of the standard Sierpinski carpet, $v = 2.194$ and $\beta = 0.234$, according to our data, while $v = 2.13$ and $\beta = 0.27$, according to the results in [19]). Other critical exponents can be found from the system of equalities for the two-exponent scaling [4]: the index of mean length of a finite cluster is $\gamma = v \cdot d - 2\beta = 3.216$, the critical exponent for the analog of specific heat is $\alpha = 2 - v \cdot d = -1.602$, and the index related to the largest size of finite clusters is $\Delta = v \cdot d - \beta = 1.809$.

In [20] investigated percolation phenomena in various multifractal objects. There are basically two sources of these differences: first is related to the coordination number, which changes along the multifractal, and second comes from the way the weight of each cell in the multifractal affects the percolation cluster. Authors use many various finite size lattices and draw the histogram of percolating lattices against site occupation probability. Authors observed that the percolation threshold for the multifractal is lower than that for the square lattice, and also computed the fractal dimension of the percolating cluster and the critical exponent beta. Despite the topological differences was shown that the percolation in a multifractal support is in the same universality class as standard percolation.

## 6. Electrical Properties and Others Applications of Theory

The review [21] had described research of the electrical characteristics after introducing carbon nanotubes into polymer matrices of composite materials. It appears 0.01-0.1% of doping is enough to increase the conductivity of the material by more than ten orders of magnitude, this changing it from an insulator to a conductor. At low doping, charge transfer is of percolation nature in the sense that nanotubes, which are in contact with each other, form conducting channels in the material. Importantly, the conductivity has a threshold nature, so that the conduction jump occurs at an arbitrarily small increase in doping above the critical value.

The review [21] summarizes experimental data about the percolation threshold and the maximum magnitude of the conductivity for composites obtained using various polymer types and various geometries of carbon nanotubes. Factors affecting the electrical characteristics of composites produced by various methods are analyzed.

It should be noted that the development of research of polymer composites doped by carbon nanotubes has an impact on the methodology of general approach to theoretical description of such systems. In particular, when used the nanotubes as a doped, which are characterized by a large aspect ratio to the forefront, becomes the problem of anisotropy of physical, chemical and electrical properties of such materials [21].

Interestingly, the rheological properties of nanocomposites also show a threshold dependence on the content of nanotubes, but the percolation threshold is lower than for electrical conductivity [22].

Macroscopic properties of heterogeneous media are frequently modeled by regular lattice models, which are based on a relatively small basic cluster of lattice sites [23]. The focus is on the percolation behavior of the effective conductivity of random two- and three-phase systems. Authors consider only the influence of geometrical features of local configurations at different length scales. At scales accessible numerically, shown that an increase in the size of the basic cluster leads to characteristic displacements of the percolation threshold. In [23] argue that the behavior is typical of materials, whose conductivity is dominated by a few linear, percolation-like, conducting paths. Such a system can be effectively treated as one-dimensional medium, and also develop a simplified model that permits an analysis at any scale. It is worth mentioning that the latter approach keeps the same thresholds predicted by the former one.

The dynamics of infiltration of a nanoporous body with a nonwetting liquid under rapid compression is studied experimentally and theoretically [24]. Experiments are carried out on systems formed by a hydrophobic nanoporous body Libersorb 23, water, and an aqueous solution of $CaCl_2$ at a compression rate of $p \geq 10^4$ atm/s. It is found that the infiltration begins and occurs at a new constant pressure independent of the compression energy and viscosity of the liquid. The time of infiltration and the filled volume increase with the compression energy. A model of infiltration of a nanoporous body with a nonwetting liquid is constructed; using this model, infiltration is described as a spatially nonuniform process with the help of distribution functions for clusters formed by pores accessible to infiltration and filled ones. On the basis of the proposed system of kinetic equations for these distribution functions, it is shown that under rapid compression, the infiltration process must occur at a constant pressure $p_c$ whose value is controlled by a new infiltration threshold $\theta_c = 0.28$ for the fraction of accessible pores, which is higher than percolation threshold $\theta_{c0} = 0.18$. Quantity $\theta_c$ is a universal characteristic of porous bodies. It is shown that the solution to the system of kinetic equations leads to a nonlinear response by the medium to an external action (rapid compression), which means the compensation of this action by percolation of the liquid from clusters of filled pores of finite size to an infinitely large cluster of accessible but unfilled pores.

The impurity transport regimes in percolation media with a finite correlation length, which are caused by advection and diffusion mechanisms, have been analyzed [25]. It has been shown that the change in the transport characteristics of a medium from the self-similar type to the statistically homogeneous type occurs through two stages because of the structural features of percolation clusters (presence of a backbone and dead ends). As a result, new anomalous transport regimes appear in the system. The quasi-isotropic and moderately and strongly anisotropic media have been considered.

## 7. Conclusions

The review absolutely not lay claim to "complete" description of the state of problems in this area of research: it's description of interesting papers published in the last 15-20 years, and it is an attempt to draw attention to some of the emerging trends in the research and interpretation influence of linked areas on the properties of materials.